\documentclass[aps,prb,twocolumn,superscriptaddress,notitlepage]{revtex4-1}
\usepackage{graphicx}
\usepackage[caption=false]{subfig}
\usepackage{float}
\usepackage{natbib}
\usepackage{xcolor}
\usepackage{xr}
\usepackage{amsmath}
\usepackage{amssymb}
\usepackage{array}
\usepackage{wasysym}
\usepackage{color,soul}
\usepackage{braket}
\usepackage{verbatim}
\usepackage{hyperref}
\usepackage{gensymb}
\usepackage{url}
\usepackage{dirtytalk}
\usepackage{upgreek}

\usepackage{filecontents}
\begin{document}

\title{Unified pressure and field response across distinct charge-order regimes in Ti-doped CsV$_3$Sb$_5$}

\author{P. Kral}
\affiliation{PSI Center for Neutron and Muon Sciences, 5232 Villigen PSI, Switzerland}

\author{S.S. Islam}
\affiliation{PSI Center for Neutron and Muon Sciences, 5232 Villigen PSI, Switzerland}

\author{Andrea N. Capa Salinas}
\affiliation{Materials Department and California Nanosystems Institute, University of California Santa Barbara, 93106 Santa Barbara, USA}

\author{J.N. Graham}
\affiliation{PSI Center for Neutron and Muon Sciences, 5232 Villigen PSI, Switzerland}

\author{O. Gerguri}
\affiliation{PSI Center for Neutron and Muon Sciences, 5232 Villigen PSI, Switzerland}
\affiliation{Physik-Institut, Universit\"{a}t Z\"{u}rich, Winterthurerstrasse 190, CH-8057 Z\"{u}rich, Switzerland}

\author{A. Doll}
\affiliation{PSI Center for Neutron and Muon Sciences, 5232 Villigen PSI, Switzerland}

\author{J. Krieger}
\affiliation{PSI Center for Neutron and Muon Sciences, 5232 Villigen PSI, Switzerland}

\author{T.J. Hicken}
\affiliation{PSI Center for Neutron and Muon Sciences, 5232 Villigen PSI, Switzerland}

\author{G. Simutis}
\affiliation{PSI Center for Neutron and Muon Sciences, 5232 Villigen PSI, Switzerland}

\author{H. Luetkens}
\affiliation{PSI Center for Neutron and Muon Sciences, 5232 Villigen PSI, Switzerland}

\author{R. Khasanov}
\affiliation{PSI Center for Neutron and Muon Sciences, 5232 Villigen PSI, Switzerland}

\author{S.D. Wilson}
\affiliation{Materials Department and California Nanosystems Institute, University of California Santa Barbara, 93106 Santa Barbara, USA}

\author{Z. Guguchia}
\email{zurab.guguchia@psi.ch} 
\affiliation{PSI Center for Neutron and Muon Sciences, 5232 Villigen PSI, Switzerland}

\date{\today}

\begin{abstract}

Understanding the phase diagram of kagome superconductors from a microscopic perspective is crucial for clarifying the interplay between charge order and superconductivity. Ti-doped CsV$_{3}$Sb$_{5}$ exhibits a nonmonotonic temperature–doping phase diagram in which both $T_{\rm c}$ and the charge-order temperature initially decrease with doping, followed by a crossover from long-range to short-range charge order and a subsequent increase in $T_{\rm c}$. Here, we report a muon spin rotation ($\mu$SR) study of Ti-doped CsV$_{3}$Sb$_{5}$ at two representative compositions: underdoped (Ti$_{0.05}$–CVS) and optimally doped (Ti$_{0.22}$–CVS). Using zero-field, high-field, and high-pressure $\mu$SR, we find spontaneous time-reversal-symmetry (TRS) breaking in the normal state of both compositions, strongly enhanced by an applied magnetic field and associated with long-range and short-range charge-order correlations, respectively. In the superconducting state, both samples exhibit anisotropic nodeless pairing with low superfluid density. Hydrostatic pressure substantially enhances both $T_{\rm c}$ and the superfluid density (by $\sim$2.5), revealing a linear correlation between them and pointing to unconventional pairing. Above $\sim$1 GPa, a crossover from anisotropic to isotropic nodeless pairing is observed. Despite the different nature of charge order in the two doping regimes, the superconducting responses are remarkably similar, suggesting that the competition between superconductivity and charge order occurs on a local scale, largely independent of the long-range coherence of the charge-ordered state.

\end{abstract}

\maketitle

\section{Introduction}
Understanding the mechanisms of unconventional superconductivity is a key objective in the study of quantum matter, particularly in materials where superconductivity emerges from a complex landscape of intertwined electronic orders. In this regard, the kagome lattice has attracted intense interest as one of the most impactful structural motives \cite{syozi1951statistics,yin2022topological,guguchia2023unconventional,wang2023quantum,di2026kagome}. Owing to its unique geometry of corner-sharing triangles, which gives rise to strong electronic frustration and a rich band structure, the kagome lattice can host a wide spectrum of emergent phenomena, including unconventional superconductivity, frustrated magnetism, and various forms of lattice instabilities manifested through charge ordering, often at elevated temperatures. Recently, charge ordering with an onset temperature of 780 K has been reported in the kagome superconductor YRu$_3$Si$_2$ \cite{kral2025discovery}, representing an unprecedentedly high ordering temperature for quantum materials. One of the most prominent realizations of kagome physics is provided by the family of \textit{A}V$_3$Sb$_5$ (\textit{A} = K, Rb, Cs), which have emerged as a model system for studying the interplay between superconductivity and charge order in a geometrically frustrated lattice \cite{ortiz2019new,mielke2022time,neupert2022charge,wilson2024v3sb5,khasanov2022time,guguchia2023tunable}. All members of this family undergo a transition into a charge-ordered state at elevated temperatures, while superconductivity develops at lower temperatures and competes for the same electronic states, establishing an intrinsically antagonistic relationship between these two collective phases. Of particular interest is the emergence of spontaneous breaking of time-reversal symmetry (TRS) \cite{mielke2022time,jiang2021unconventional,graham2024depth,khasanov2022time,huang2026magnetic,Guo2022,gui2025probing,xing2024optical,Yu2021arxiv} in the charge ordered state, first reported for KV$_3$Sb$_5$ \cite{mielke2022time,jiang2021unconventional}, highlighting the kagome lattice as a fertile platform for studying correlated and topologically nontrivial electronic states.

\begin{figure*}
    \centering
    \includegraphics[width=1.0\linewidth]{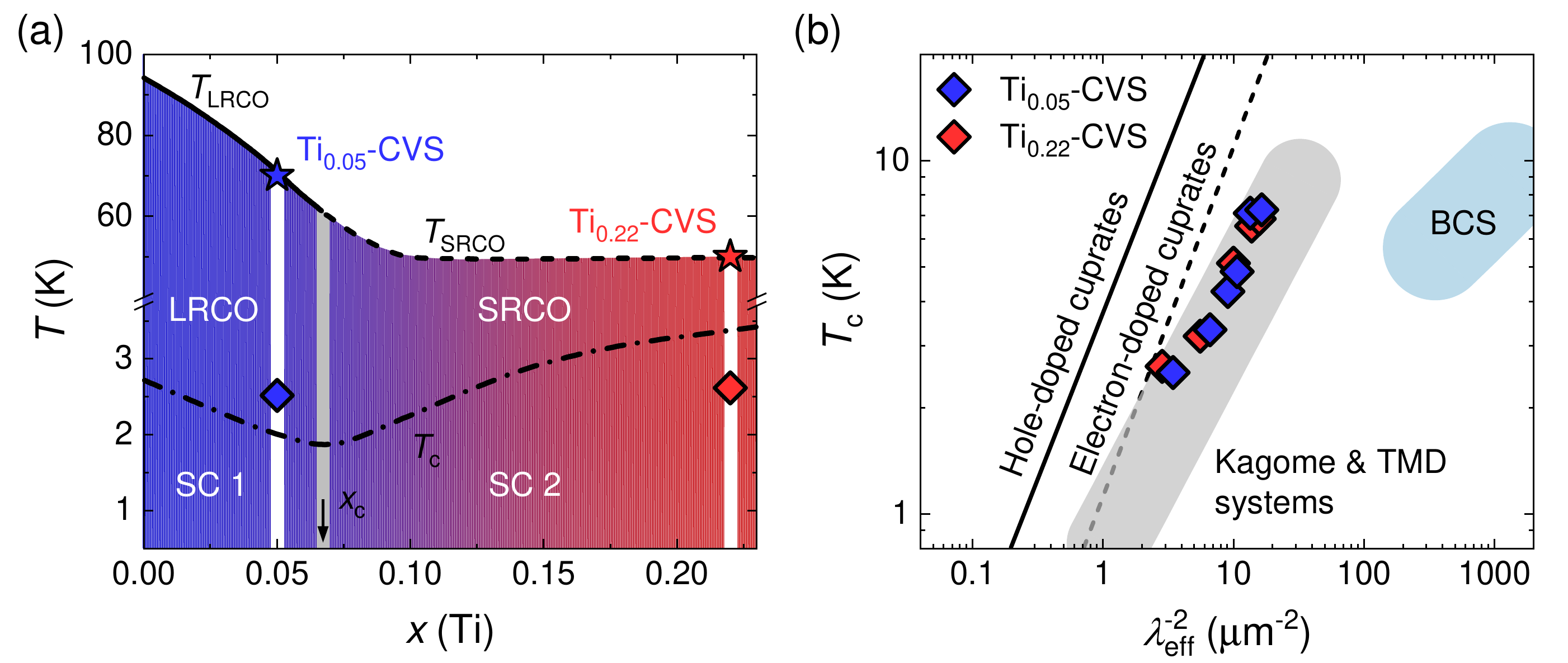}
    \caption{\textbf{Schematic phase diagram and superconducting scaling in Ti-doped CsV$_3$Sb$_5$. a} Electronic phase diagram of Ti-doped CsV$_3$Sb$_5$, illustrating the Ti-concentration dependence of charge order and superconductivity. Color gradient indicates the transition between long-range (LRCO) and short-range (SRCO) charge order and corresponding superconducting regimes SC 1 and SC 2. Vertical gray line corresponds to the crossover doping level. Black lines represent the schematic Ti-concentration dependencies of charge-order transition temperature (solid for long-range ($T_{\rm{LRCO}}$) and dashed for short-range ($T_{\rm{SRCO}}$) regimes, respectively). The dashed-dotted schematic  line for the superconducting transition temperature $T_{\rm{c}}$ is taken as an average through a large number of data points reported in the literature \cite{yang2022titanium,hou2023effect,wu2023unidirectional,wu2025competitive,pokharel2025evolution}. Vertical white lines indicate two particular doping levels, ${x=0.05}$ and ${x=0.22}$ investigated in this study. The transition temperatures determined from our ${\mu}$SR experiments are marked by star (CO) and diamond (SC) symbols, in blue and red colors for Ti${_{0.05}}$-CVS and Ti${_{0.22}}$-CVS, respectively. \textbf{b} Plot of \(T_\text{c}\) versus \(\lambda_\text{eff}^{-2}(0)\) in logarithmic scale summarizing results of our ${\mu}$SR experiments as a function of pressure. Dependencies reported previously for hole- and electron-doped cuprates\cite{uemura1989universal,uemura1991basic,uemura2004condensation,uemura2009energy,shengelaya2005muon} and various kagome\cite{mielke2022time,guguchia2023tunable, gupta2022microscopic,mielke2021nodeless,mielke2022local} and TMD\cite{guguchia2017signatures,von2019unconventional, sazgari2025competing} materials (gray shaded area) are shown for comparison, as well as the region of typical BCS values.}
    \label{fig:PD}
\end{figure*}

Importantly, external tuning parameters such as hydrostatic pressure and chemical substitution have been shown to effectively suppress the charge-ordered state across the \textit{A}V$_3$Sb$_5$ family, leading to a pronounced enhancement of the superconducting transition temperature \cite{guguchia2023tunable}. Remarkably, when the charge order is fully suppressed, the superconducting state itself has been reported \cite{guguchia2023unconventional} to spontaneously break time-reversal symmetry---a rare and highly unconventional property that underscores the nontrivial nature of superconductivity in kagome metals. Among the three compounds, CsV$_3$Sb$_5$ stands out due to its comparatively high superconducting transition temperature (\(T_\text{c}\) $\simeq$ 2.5 K) and the complex interplay between superconductivity and charge order, where two types of charge order give rise to a double-dome structure in the superconducting phase under pressure before the charge order is fully suppressed and the superconducting transition temperature jumps to 8 K\cite{gupta2022two,zheng2022emergent}. Beyond pressure, chemical substitution provides an alternative and highly effective route to tune the balance between charge order and superconductivity in CsV$_3$Sb$_5$. Partial replacement of the V kagome site with Nb or Ta has been shown to suppress the charge-ordered state, leading to an enhancement of the superconducting transition temperature \cite{zhong2023nodeless,graham2024depth,liu2024enhancement,graham2025pressure}. When combined with additional control parameters---such as magnetic field, depth profiling, and hydrostatic pressure---this strategy offers a powerful way to explore the interplay of these intertwined phases, as demonstrated in our recent work on Nb-doped CsV$_3$Sb$_5$ \cite{graham2025pressure}.

Recently, Ti-doped CsV$_3$Sb$_5$ has emerged as a particularly intriguing variant of \textit{A}V$_3$Sb$_5$ \cite{yang2022titanium,sur2023optimized,hou2023effect,wu2023unidirectional,liu2023doping,huang2024tunable,huai2024two,wu2025competitive,pokharel2025evolution,shtefiienko2025pressure,huang2025revealing,sun2025intertwined,luo2025van,shtefiienko2026haas,xiao2026evolution}. With increasing Ti concentration, the superconducting transition temperature is initially suppressed---from 2.5 K in the parent compound to ${\approx}$2 K at ${x=0.07}$ doping. At higher doping levels, however, the system undergoes a qualitative change in its superconducting state, accompanied by a subsequent enhancement of \(T_\text{c}\) to nearly 3.5 K at the optimal composition ${x=0.22}$. Notably, this evolution coincides with the doping level at which charge order collapses into the short-range regime with the onset temperature of $\simeq$55 K\cite{xiao2026evolution}, suggesting a strong link between the loss of charge order spatial coherence and the emergence of the enhanced superconducting phase (Fig. \ref{fig:PD}a). This behavior contrasts with the phase diagrams of other dopants: here, \(T_\text{c}\) initially decreases as \(T_\text{co}\) is reduced, implying cooperation rather than competition between superconductivity and charge order. Once the long-range charge order is fully suppressed, however, \(T_\text{c}\) begins to rise. A key question is whether the superconducting states below and above the crossover doping ${x=0.07}$ share the same characteristics or are fundamentally different. It is therefore essential to probe the microscopic details of both the superconducting and normal-state properties in samples from these two regions of the phase diagram.

In this article, we report insights into the microscopic behavior as revealed by muon spin rotation/relaxation (${\mu}$SR) in both the normal and superconducting states of two specific compositions of CsV$_{3-x}$Ti$_{x}$Sb$_{5}$ systems---the underdoped ${x=0.05}$ (Ti${_{0.05}}$-CVS) and optimally doped ${x=0.22}$ (Ti${_{0.22}}$-CVS). Combined zero-field, high-field, and high-pressure ${\mu}$SR experiments reveal three key findings common to both compositions, as follows: (1) spontaneous time-reversal symmetry breaking in the normal state, associated with long-range and short-range charge-order correlations in the underdoped and optimally doped regimes, respectively; (2) a pronounced pressure-induced enhancement of both the superconducting transition temperature and the superfluid density, establishing a linear correlation between these quantities—a hallmark of unconventional superconductivity; and (3) a pressure-driven crossover from anisotropic nodeless to isotropic nodeless gap symmetry.
Despite the fundamentally different nature of long-range and short-range charge-order correlations across the two doping regimes, the superconducting responses remain remarkably similar. This indicates that the competition between superconductivity and charge order is governed by local electronic correlations, largely independent of the long-range coherence of the charge-ordered state.

\section{Results}
\subsection{Normal-state properties}

We begin our investigation by examining the normal-state properties of Ti-doped CsV$_3$Sb$_5$ for two representative compositions: the underdoped Ti${_{0.05}}$-CVS, which retains charge order, and the optimally doped Ti${_{0.22}}$-CVS, in which long-range charge order is suppressed but short-range charge order persists.
To probe the microscopic differences between these two normal states, we employ muon spin rotation/relaxation (${\mu}$SR), a local probe technique capable of detecting weak internal magnetic fields as small as 0.01 mT, without the need for the application of an external magnetic field. This sensitivity makes ${\mu}$SR particularly well suited for detecting spontaneous internal magnetic fields associated with broken time-reversal symmetry.

\begin{figure*}
    \centering
    \includegraphics[width=1.0\linewidth]{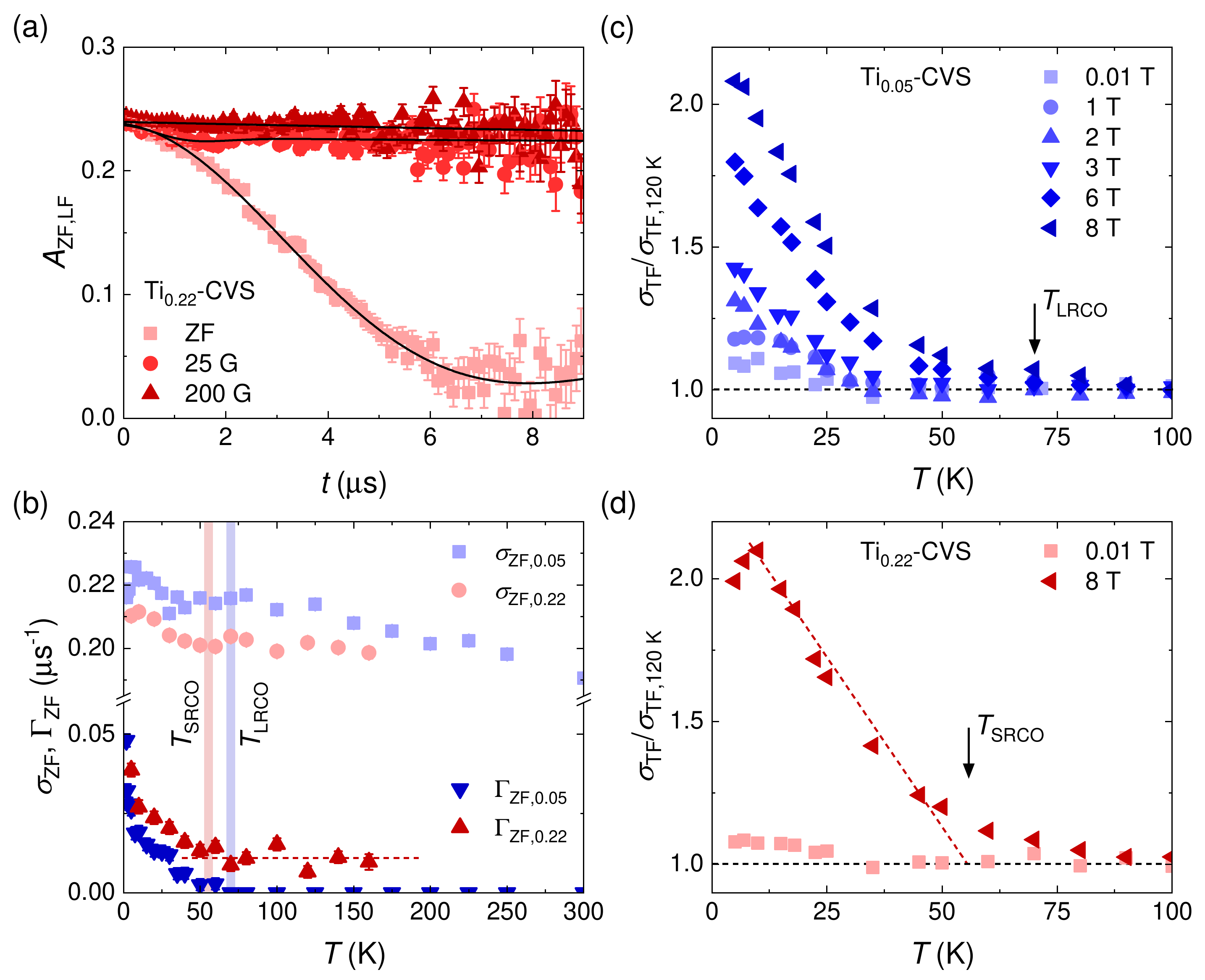}
    \caption{\textbf{Normal-state properties of Ti-doped CsV$_3$Sb$_5$ probed by ${\mu}$SR. a} Zero-field and longitudinal-field ${\mu}$SR spectra recorded at 5 K for Ti${_{0.22}}$-CVS. The black solid lines represent fits using Eq. \ref{eqGKT2}.  Error bars are the standard error of the mean (s.e.m.) in about ${10^6}$ events. \textbf{b} Temperature dependence of the relaxation rates ${\sigma}$ (light) and ${\Gamma}$ (dark) obtained for Ti${_{0.05}}$-CVS (blue) and Ti${_{0.22}}$-CVS (red), respectively. Vertical lines in corresponding colors mark the long-range and short-range charge-order transition temperatures \(T_\text{LRCO}\)\cite{wu2025competitive} and \(T_\text{SRCO}\)\cite{xiao2026evolution}, determined in previous works \cite{wu2025competitive,xiao2026evolution}. \textbf{c,d} Temperature dependence of the high transverse-field muon spin relaxation rates measured in selected external fields for Ti${_{0.05}}$-CVS (\textbf{c}) and Ti${_{0.22}}$-CVS (\textbf{d}), respectively.} 
    \label{fig:NS}
\end{figure*}

In Fig. \ref{fig:NS}a,b, we first summarize the zero-field (ZF) ${\mu}$SR measurements for both samples. Fig. \ref{fig:NS}a presents the ${\mu}$SR time spectra of the Ti$_{0.22}$-CVS sample recorded at 5 K in ZF and longitudinal-field (LF, field applied parallel to the muon spin polarization) configurations. The full polarization can be recovered by the application of a small external longitudinal magnetic field of 20 mT. This indicates that the observed relaxation arises from spontaneous internal fields that are static on the microsecond timescale. The ZF ${\mu}$SR data were fitted using a Gaussian Kubo-Toyabe (GKT) depolarization function multiplied by an exponential relaxation term (Eq. \ref{eqGKT2}, for details see the "Methods" section), introducing two characteristic parameters, $\sigma$ and $\Gamma$. The Gaussian relaxation rate $\sigma$ represents the width of the local field distribution primarily due to the nuclear moments. However, a dense distribution of electronic moments can also contribute to the relaxation rate.
The exponential relaxation rate $\Gamma$ is attributed predominantly to electronic moments, as supported by the field dependence discussed below.

The temperature dependencies of the relaxation rates $\sigma$ and $\Gamma$ are shown in Fig.~\ref{fig:NS}b. While $\sigma$ exhibits only a subtle non-monotonic variation across $T_\text{co}\simeq 70$ K in Ti$_{0.05}$-CVS, followed by a modest increase below $\approx$30 K in both systems, a much more pronounced effect is observed in $\Gamma$. In Ti$_{0.05}$-CVS, a significant enhancement of $\Gamma$ sets in below the long-range charge-order temperature $T_\text{LRCO}\simeq 70$ K. A similar increase is detected in Ti$_{0.22}$-CVS below the short-range charge-order temperature $T_\text{SRCO}\simeq 55$ K. At 2 K, the enhancement in $\Gamma$ amounts to approximately 0.05 $\upmu$s$^{-1}$ for Ti$_{0.05}$-CVS and 0.03 $\upmu$s$^{-1}$ for Ti$_{0.22}$-CVS, corresponding to characteristic internal fields of about 0.3–0.6 G. Notably, $\Gamma$ exhibits an additional upturn at the lowest few temperatures points, particularly pronounced in Ti$_{0.05}$-CVS, resulting in a two-step increase of the relaxation rate. A similar two-step temperature evolution has been reported in the undoped \textit{A}V$_3$Sb$_5$ systems \cite{khasanov2022time,guguchia2023tunable,bonfa2025unveiling}. Although the precise origin of this behavior remains unclear, its recurring observation across various 135 kagome systems is intriguing and suggests a potentially generic underlying mechanism.

To further substantiate the magnetic origin of the ZF ${\mu}$SR signal, complementary transverse-field (TF) ${\mu}$SR measurements were carried out in magnetic fields of up to 8 T. In the TF geometry, distinct relaxation channels cannot be reliably disentangled. The spectra were therefore described using a single effective muon spin relaxation rate, $\sigma_{\rm{TF}}$. The temperature dependence of $\sigma_{\rm{TF}}$ under different applied fields is shown in Fig. \ref{fig:NS}c and d, for Ti${_{0.05}}$-CVS and Ti${_{0.22}}$-CVS, respectively. At the lowest applied field of 0.01 T, only a weak increase is observed below $\approx$30 K, closely resembling the behavior of $\sigma$ at the zero field. Upon increasing the applied field, the enhancement becomes significantly stronger with the onset shifted towards higher temperatures, reflecting the dominant contribution of $\Gamma$ rate. The absolute magnitude of the increase grows monotonically with field strength with more than five-fold enhancement at the highest applied field of 8 T. Because an external magnetic field does not amplify nuclear relaxation and, in high fields, the muon quantization axis is governed by the applied field rather than the electric field gradient\cite{mielke2022time}, the observed field-induced enhancement at low temperatures indicates a dominant electronic (magnetic) origin of the relaxation. Therefore, the combined use of zero-field and high-field ${\mu}$SR measurements is essential to establish the presence of time-reversal symmetry breaking in the normal state of these superconductors.

\subsection{Superconducting properties}

\begin{figure*}
    \centering
    \includegraphics[width=1.0\linewidth]{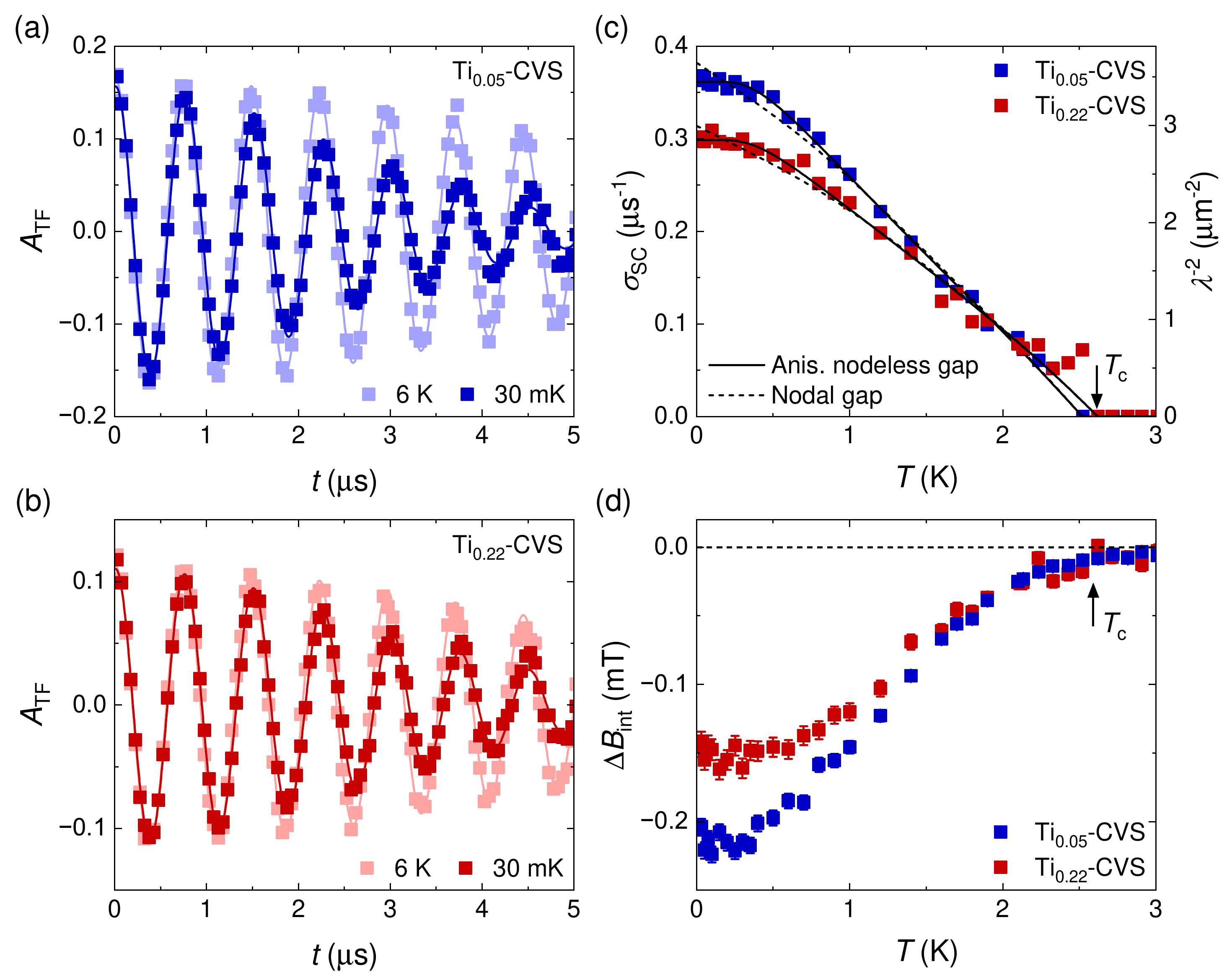}
    \caption{\textbf{Superconducting-state properties of Ti-doped CsV$_3$Sb$_5$ probed by ${\mu}$SR. a,b} Transverse field (TF) ${\mu}$SR spectra collected above (6 K) and below (30 mK) the superconducting transition after field-cooling from above \(T_\text{c}\) in an applied field of 10 mT, for Ti${_{0.05}}$-CVS and Ti${_{0.22}}$-CVS, respectively. Error bars are the standard error of the mean (s.e.m.) in about ${10^6}$ events. \textbf{c} Temperature dependence of the superconducting muon spin depolarization rate, \(\sigma_\text{SC}\) (left axis) and inverse squared effective penetration depth, \(\lambda_\text{eff}^{-2}(T)\) (right axis) measured in 10 mT, together with the fits to theoretical models. The error bars represent the standard error extracted from the covariance matrix in the fitting. \textbf{d} Diamagnetic shift $\Delta B_\text{int}$ at 10 mT.}
    \label{fig:SC}
\end{figure*}

Having established the normal-state properties of both systems, we now turn to the superconducting state, explored microscopically by transverse-field (TF) ${\mu}$SR measurements at ambient conditions and under applied hydrostatic pressure. A summary of the ambient-pressure superconducting properties is shown in Fig. \ref{fig:SC}. Namely, Fig. \ref{fig:SC}a and b show the TF $\mu$SR time spectra measured in an applied magnetic field of 10 mT above (6 K) and below (30 mK) the superconducting transition, for Ti${_{0.05}}$-CVS and Ti${_{0.22}}$-CVS, respectively. Above \(T_\text{c}\), the oscillations are weakly damped due to the random local fields from the nuclear moments, whereas below \(T_\text{c}\) the damping rate increases significantly, reflecting the inhomogeneous local field distribution associated with the formation of a flux-line lattice. In the superconducting state, the corresponding internal field distribution becomes broadened and asymmetric, with the first moment shifted below the applied field, allowing to determine the superconducting diamagnetic shift ($\Delta B_{\mathrm{int}} = \mu_0 ( H_{\mathrm{int,SC}} - H_{\mathrm{int,NS}} )$). The data presented in Fig. \ref{fig:SC}d show a clear decrease in $\Delta B_{\mathrm{int}}$, further confirming the bulk nature of superconductivity. For Ti${_{0.05}}$-CVS, a slightly larger value of 0.21 mT is obtained, whereas Ti${_{0.22}}$-CVS exhibits a reduced shift of 0.15 mT. These magnitudes are comparable to those reported for the undoped AV$_3$Sb$_5$ compounds CsV$_3$Sb$_5$\cite{gupta2022microscopic} and KV$_3$Sb$_5$\cite{mielke2022time}, as well as for Nb-doped CsV$_3$Sb$_5$\cite{graham2025pressure}.

Detailed insights into the superconducting properties can be gained from the analysis of the muon spin depolarization rate, \(\sigma_\text{tot}\) (\(=\sqrt{\sigma_\text{SC}^2+\sigma_\text{nm}^2}\)), consisting of superconducting, \(\sigma_\text{SC}\) and nuclear magnetic dipolar, \(\sigma_\text{nm}\) contributions. Quadrature subtraction of the nuclear contribution, assuming it remains constant above \(T_\text{c}\), allows for the estimation of the superconducting relaxation rate \(\sigma_\text{SC}\).
The temperature dependencies of \(\sigma_\text{SC}\) for both Ti${_{0.05}}$-CVS and Ti${_{0.22}}$-CVS are shown in Fig. \ref{fig:SC}c. For a perfectly ordered triangular vortex lattice, \(\sigma_\text{SC}\) is proportional to the inverse square of the magnetic penetration depth, $\lambda_{\text{eff}}^{-2}$ (right axis in Fig. \ref{fig:SC}c), providing one of the most fundamental parameters characterizing a superconductor. The magnetic penetration depth is related to the superfluid density as 1/${\lambda}^{2}$=${\mu}_{0}$$e^{2}$$n_{\rm s}$/$m^{*}$ (where $m^*$ is the effective mass), thereby offering a direct measure of the number of Cooper pairs. These results confirm the bulk superconductivity with \(T_\text{c}=2.52(3)\) K and \(T_\text{c}=2.62(3)\) K for Ti${_{0.05}}$-CVS and Ti${_{0.22}}$-CVS, respectively. Moreover, consistent with other kagome materials, both systems exhibit low value of the superfluid density, with \(T_\text{c}/\lambda_\text{eff}(0)^{-2}\) values of 0.73 K$\upmu$m$^{2}$ and 0.92 K$\upmu$m$^{2}$, placing both systems close to the regime of unconventional superconductivity (see Fig. \ref{fig:PD}b). The extracted zero-temperature effective penetration depths are $\lambda_\text{eff}(0)=538(3)$ nm and $\lambda_\text{eff}(0)=592(3)$ nm for Ti$_{0.05}$-CVS and Ti$_{0.22}$-CVS, respectively.

\begin{figure*}
    \centering
    \includegraphics[width=1.0\linewidth]{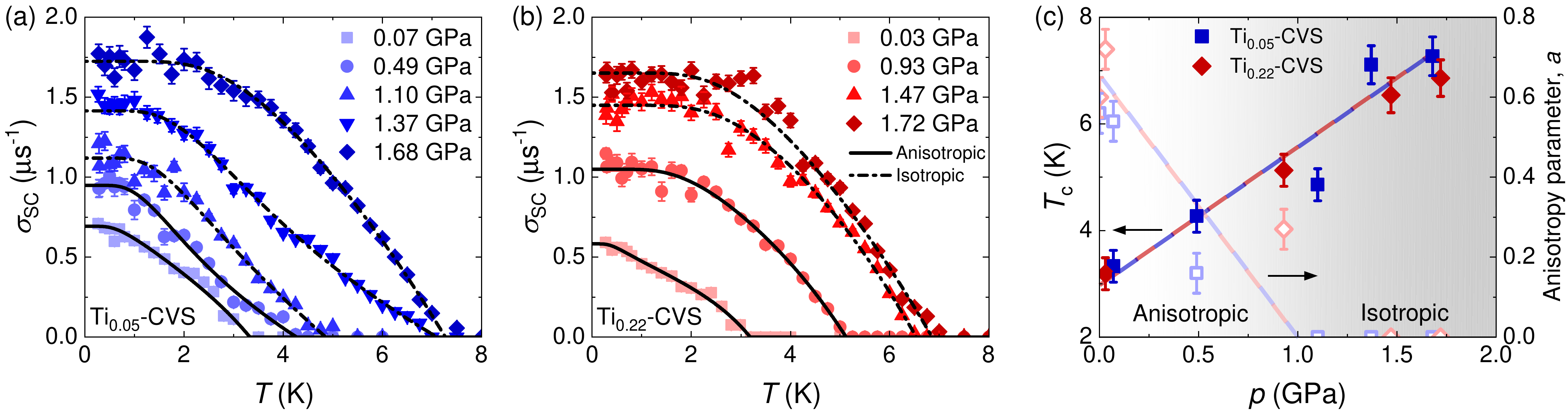}
    \caption{\textbf{Summary of the pressure impact on superconducting properties of Ti-doped CsV$_3$Sb$_5$ as revealed by ${\mu}$SR measurements.  a,b} Superconducting muon spin depolarization rates of Ti${_{0.05}}$-CVS and Ti${_{0.22}}$-CVS, respectively, as functions of pressure, together with the fits to theoretical models. Solid lines correspond to the anisotropic nodeless model, transitioning to the isotropic nodeless gap structure at elevated pressures (dashed-dotted lines). \textbf{c} Pressure phase diagram summarizing the superconducting properties of both systems, showing the pressure dependencies of the superconducting transition temperature \(T_\text{c}\) (left axis) and the anisotropy parameter of the superconducting gap (right axis). Background gradient indicates the transition between anisotropic and isotropic nodeless gap structure.}
    \label{fig:P}
\end{figure*}

For a quantitative assessment, the experimental data were analyzed using several commonly used gap models. In both systems, the temperature dependence of superfluid density shows a tendency toward saturation at low temperatures, indicative of a nodeless superconducting gap structure. However, the onset of saturation occurs at temperatures significantly lower than the \(T_\text{c}/3\) scale expected for an isotropic nodeless $s$-wave gap. Among the considered models, the anisotropic nodeless superconducting gap provides the most consistent description of the experimental data. Fig. \ref{fig:SC}c compares the fits obtained using the anisotropic $s$-wave and nodal $d$-wave models, with the former providing a substantially better description of low-temperature behavior and thus favoring a nodeless gap structure.
The angular dependence of the gap can be parametrized as $\Delta(\varphi)=\Delta_0(1+a\cos(4\varphi))$. The extracted gap amplitudes are $\Delta_0 = 0.28(1)$ meV and $0.31(1)$ meV for Ti$_{0.05}$-CVS and Ti$_{0.22}$-CVS, respectively, with corresponding anisotropy parameters $a = 0.56(2)$ and $0.59(2)$. The large values of the anisotropy parameters indicate a strongly anisotropic superconducting gap structure.

Next, we focus on the response of the superconducting state to hydrostatic pressure. In kagome superconductors, and in unconventional superconductors more generally, hydrostatic pressure has proven to be a particularly effective tuning parameter, capable of significantly modifying the superconducting properties. In unconventional superconductors, the pressure response of superconductivity is often intimately linked to the pressure evolution of other electronic orders that are established at higher temperatures. In kagome superconductors, such a correlated phase is most often realized in the form of charge order. The relationship between charge order and superconductivity, however, is not universal and can be either cooperative or antagonistic, depending on the material. While the positive intertwining is relatively uncommon (with LaRu$_3$Si$_2$ being a rare example \cite{ma2025correlation}), in many other systems---including \textit{A}V$_3$Sb$_5$ family---superconductivity is generally enhanced as charge order is weakened or suppressed \cite{guguchia2023tunable}. Combining the two main experimental strategies allowing for this effect, i.e., applying mechanical pressure on chemically doped material, offers a particularly interesting direction for exploring the underlying mechanisms.

In Fig. \ref{fig:P}a,b, we show the effect of hydrostatic pressure on the temperature dependence of the superconducting relaxation rate \(\sigma_\text{SC}\) for Ti${_{0.05}}$-CVS and Ti${_{0.22}}$-CVS, respectively. In both systems, \(T_\text{c}\) increases continuously with pressure and, intriguingly, saturates at a similar maximum value of about 7 K. Concomitantly, the superfluid density is enhanced by nearly a factor of three, demonstrating a clear linear scaling relationship between \(T_\text{c}\) and the superfluid density. The results are summarized in the phase diagram, Fig. \ref{fig:P}c, showing the pressure dependence of \(T_\text{c}\).
In the underdoped Ti${_{0.05}}$-CVS, the pressure-induced increase in both \(T_\text{c}\) and superfluid density naturally aligns with the scenario of competition between superconductivity and long-range charge order, progressively weakened under pressure. Intriguingly, a comparable enhancement in Ti${_{0.22}}$-CVS suggests---in accordance with the persistence of short-range charge-order correlations\cite{xiao2026evolution}---that the optimally doped sample also hosts some competing order that is suppressed under pressure. This result provides additional confirmation of presence of the electronic phase emerging below 50 K in Ti${_{0.22}}$-CVS, as indicated by the ZF and TF ${\mu}$SR experiments discussed above.

In addition to the substantial enhancement of both \(T_\text{c}\) and the superfluid density, pressure induces another notable change: a modification of the superconducting gap structure in both materials. Given the anisotropic nodeless superconductivity identified at ambient pressure, the same anisotropic $s$-wave model was initially employed to describe the data at elevated pressures. For pressures below 1 GPa, this model continues to provide an excellent description of the experimental results. However, with further increasing pressure, the gap anisotropy progressively diminishes, as reflected by the anisotropy parameter smoothly approaching zero (Fig. \ref{fig:P}c, right axis). In this regime, the superconducting state is effectively described by an isotropic nodeless gap.

\section{Discussion}

Taken together, the results presented above provide a comprehensive microscopic study of both the superconducting and normal-state properties of Ti-doped CsV$_3$Sb$_5$, spanning regions of the phase diagram characterized by long-range and short-range charge order. Below, we summarize the principal conclusions of this work.

(1) The combination of zero-field (ZF) and high transverse-field (TF) $\mu$SR measurements provides direct evidence for time-reversal symmetry (TRS) breaking associated with charge order. In underdoped Ti$_{0.05}$-CVS, TRS breaking emerges at the onset of long-range charge order ($T_\text{LRCO}\simeq 70$ K), while in optimally doped Ti$_{0.22}$-CVS it appears in conjunction with short-range charge order ($T_\text{SRCO}\simeq 55$ K). Remarkably, both compositions exhibit a comparable absolute field-induced enhancement of the relaxation rate, indicating that TRS breaking is driven by local charge-order correlations and does not require long-range coherence. The increase in $\sigma_{\rm TF}$ scales linearly with applied field for both systems, in contrast to the non-linear behavior previously reported for Nb-doped CsV$_3$Sb$_5$ \cite{graham2025pressure} and for KV$_3$Sb$_5$ \cite{mielke2022time}. The magnitude of the field-induced broadening is comparable to that observed in KV$_3$Sb$_5$ but significantly larger than in undoped and Nb-doped CsV$_3$Sb$_5$ \cite{khasanov2022time,graham2025pressure}. Overall, ZF and high-TF $\mu$SR reveal a pronounced broadening of the internal field distribution, providing evidence for intrinsic TRS-breaking fields within the kagome lattice and suggesting that this phenomenon is a generic feature across the 135 kagome family.

(2) Both Ti$_{0.05}$-CVS and Ti$_{0.22}$-CVS exhibit bulk superconductivity below $T_\text{c}=2.52(3)$ K and $2.62(3)$ K, respectively. In both cases, the superconducting state is characterized by an anisotropic nodeless gap structure and an unusually low superfluid density, placing these systems in the regime of dilute superfluid superconductors. The observation of a strongly anisotropic superconducting gap together with a dilute superfluid density is consistent with an unconventional pairing mechanism. Anisotropic multigap superconductivity—comprising one isotropic and one strongly anisotropic gap—has previously been reported for undoped CsV$_3$Sb$_5$ \cite{gupta2022microscopic,zhao2025strongly}, while even more pronounced nodal gap behavior has been observed in KV$_3$Sb$_5$ and RbV$_3$Sb$_5$ \cite{guguchia2023tunable}. More generally, in undoped \textit{A}V$_3$Sb$_5$ compounds, as well as in doped systems that retain charge order, the superconducting gap structure is anisotropic and, in extreme cases, exhibits nodal behavior. A weakly anisotropic superconducting gap has also been reported in the kagome superconductor YRu$_3$Si$_2$ \cite{kral2025discovery}, a member of the 132 family, as well as in CeRu$_2$ \cite{gerguri2025distinct}, a three-dimensional structural analog of kagome systems. Together, these observations point toward a broader tendency for anisotropic pairing in kagome and kagome-related superconductors.

(3) High-pressure measurements reveal a substantial enhancement of both $T_\text{c}$ and the superfluid density in both compositions. The enhancement is comparable in magnitude for the two doping levels and results in a clear linear correlation between $T_\text{c}$ and the superfluid density. The ratio between superfluid density and $T_\text{c}$ is consistent with previously reported values for other 135 kagome systems and is characteristic of unconventional superconductors. This linear scaling represents a hallmark of unconventional pairing. The pronounced pressure-induced enhancement, particularly in the optimally doped sample, implies the presence of a competing electronic phase that is progressively weakened under pressure. Since charge order is not fully suppressed but instead evolves into short-range correlations, the observed behavior supports a scenario in which superconductivity competes locally with charge order across the phase diagram. In addition, increasing pressure gradually reduces the anisotropy of the superconducting gap, leading to a fully isotropic nodeless state above approximately 1 GPa. This demonstrates that both long-range and short-range charge order influence the superconducting gap structure. Similar pressure-induced crossovers have been observed in undoped KV$_3$Sb$_5$ and RbV$_3$Sb$_5$, although in those systems the transition occurs from nodal to nodeless behavior. Across all studied compounds—undoped and doped alike—once charge order is sufficiently suppressed by pressure, a robust nodeless superconducting state emerges. This suggests that in the absence of competing charge order, the isotropic nodeless state represents the most stable superconducting ground state, whereas gap anisotropy or nodal features arise from competition between charge order and superconductivity.

\section{Conclusion}

In conclusion, our comprehensive ${\mu}$SR study under ambient and applied pressure establishes a unified microscopic framework for superconductivity across distinct charge-order regimes in CsV$_{3-x}$Ti$_{x}$Sb$_{5}$. We demonstrate that time-reversal symmetry breaking persists in the normal state irrespective of whether charge order is long-range or short-range, highlighting the intrinsic nature of electronic symmetry breaking in this kagome system. Pressure robustly amplifies superconductivity in both regimes, producing a linear scaling between the superconducting transition temperature and superfluid density—a hallmark feature of unconventional pairing. The observation of a pressure-driven crossover from anisotropic to isotropic nodeless superconductivity further underscores the tunable and nontrivial character of the pairing state.
Strikingly, the near-identical superconducting response across fundamentally different charge-order landscapes reveals that the essential competition between superconductivity and charge order is governed by local electronic correlations rather than long-range coherence. These results provide new insight into the microscopic mechanisms operative in kagome metals and position CsV$_{3-x}$Ti$_{x}$Sb$_{5}$ as a model platform for exploring intertwined order and unconventional superconductivity in geometrically frustrated systems.

\section*{Methods}
\textbf{Sample preparation:} 
Polycrystalline samples of  CsV$_{3-x}$Ti$_{x}$Sb$_{5}$ with nominal compositions $x=0.05$ and $x=0.22$ were prepared by measuring stoichiometric amounts of elemental Cs (solid, Alfa 99.98${\%}$), V (powder, Alfa 99.9${\%}$, pre-cleaned with a 1:10 EtOH and HCl mixture), Ti (powder, Alfa 99.9${\%}$), and Sb (shot, Alfa 99.999${\%}$) inside an argon filled glovebox (oxygen and water levels at ${\textless}$0.5 ppm). The starting materials were loaded into a tungsten carbide vial and ball-milled inside a SPEX 8000D high-energy mill for 60 min. The resulting powders were extracted inside a glovebox, ground and sieved through a 
106-micron sieve, to later be placed inside alumina crucibles and sealed quartz tubes for annealing at 550$^{\circ}$C for 48 h. A second anneal at 450$^{\circ}$C for 12 h followed the powder extraction and second ground. The final powders were gray and reasonably air stable. The effective doping concentrations of the powders matched their nominal compositions, as confirmed with a tabletop scanning electron microscope (Hitachi TM4000Plus) using energy dispersive spectroscopy.\\

\textbf{Muon spin rotation/relaxation (${\mu}$SR):} In a ${\mu}$SR experiment, nearly 100${\%}$ spin-polarized positive muons $\mu$$^{+}$ are implanted into the sample one at a time. After implantation, the $\mu$$^{+}$ thermalize at interstitial lattice sites, where they act as local magnetic microprobes. In a magnetic environment, the muon spin precesses in the local field $B_{\rm \mu}$ at the Larmor frequency ${\nu}_{\rm \mu}$ = $\gamma_{\rm \mu}$/(2${\pi})$$B_{\rm \mu}$, where $\gamma_{\rm \mu}$/(2${\pi}$) = 135.5 MHz T$^{-1}$ is the muon gyromagnetic ratio. The muon subsequently decays with a mean life time of 2.2 $\upmu$s into a positron, which is preferentially emitted in the direction of the muon spin.

To microscopically probe both the superconducting and normal-state properties, a comprehensive set of ${\mu}$SR experiments was performed at the Swiss Muon Source (S${\mu}$S), Paul Scherrer Institute, Villigen, Switzerland.

Zero-field (ZF) and longitudinal-field (LF) measurements, used to detect spontaneous magnetic fields in the normal state and to assess their dynamics, were carried out on the GPS instrument. A continuous flow cryostat was used to cover the temperature range of 1.5–300 K. To extend the temperature range to well below $T_{\rm c}$ and to search for possible magnetic signatures associated with time-reversal symmetry breaking across the superconducting transition, complementary low-temperature (down to 30 mK) ZF ${\mu}$SR measurements were performed on the FLAME instrument.

To confirm the magnetic origin of the ZF ${\mu}$SR signal, additional transverse-field (TF) experiments were carried out on the HAL-9500 instrument in external fields of 0.01-8 T and over a temperature range of 5–300 K using the continuous flow cryostat Janis.

Weak-TF measurements, providing the key superconducting length scales---namely the magnetic penetration depth $\lambda$ and the coherence length $\xi$---were performed on the FLAME instrument in the magnetic field of 10 mT. Using the KelvinoxJT dilution refrigerator insert allowed us to probe the superconducting state down to the lowest temperature of 30 mK. In a type-II superconductor cooled below $T_{\rm c}$ in an applied magnetic field between the lower ($H_{c1}$) and upper ($H_{c2}$) critical fields, a vortex lattice is formed. This lattice is generally incommensurate with the crystal structure, with vortex cores separated by distances much larger than the unit-cell dimensions. Implanted muons, which stop at interstitial sites, randomly sample the local magnetic field distribution associated with the vortices.

High-pressure weak-TF measurements were conducted on the GPD instrument. Polycrystalline powder was loaded into the low-background double-wall CuBe/CuBe pressure cell with an inner diameter of 6 mm and a sample space height of 12 mm. Daphne 7373 oil was used as a pressure-transmitting medium to ensure hydrostatic conditions. The pressure value was determined by monitoring the superconducting transition of a Sn piece loaded together with the sample inside the pressure cell, using AC magnetic susceptibility measurements performed in an in-house Janis cryostat.\\

\textbf{Analysis of the ${\mu}$SR data:} The ZF ${\mu}$SR data were fitted using a Gaussian Kubo-Toyabe (GKT) depolarization function multiplied by an exponential relaxation term.

\begin{equation}
\begin{split}
	\begin{aligned}
		A_{\rm ZF}^{\rm GKT} (t) = \Big(\frac{1}{3}+\frac{2}{3}(1-\sigma^2t^2)\exp\Big[-\frac{\sigma^2t^2}{2}\Big]\Big)  \\ 
        \times \exp(-\Gamma t) + A_{\rm bkg}  
		\label{eqGKT2}
	\end{aligned}
    \end{split}
\end{equation}

Here, ${\sigma/\gamma_\mu}$ represents the width of the local magnetic field distribution arising from nuclear moments. The exponential relaxation rate $\Gamma$ primarily reflects the temperature evolution of the electronic contribution to the muon spin relaxation.

To describe the field distribution ($P (B)$) in the superconducting state, the TF ${\mu}$SR time spectra collected below $T_{\rm c}$ were analyzed using the following functional form:

\begin{equation}
	\begin{aligned}
		A_{\rm TF} (t) = A_{\rm s}\exp\Big[-\frac{\sigma^2t^2}{2}\Big]\cos(\gamma_{\mu}B_{{\rm int},\rm {s}}t+\phi)  
		\label{eqS1}
	\end{aligned}
\end{equation}

Here $A_{s}$, $\sigma = \sqrt{\sigma^{2}_\mathrm{SC} + \sigma^{2}_\mathrm{ns}}$ and $B_{{\rm int},\rm {s}}$ denote the initial asymmetry, relaxation rate, and local internal magnetic field, respectively, ${\phi}$ is the initial phase of the muon-spin ensemble.

For the high-pressure experiment, the additional contribution of the pressure cell\cite{khasanov2016high} is modeled by the same function (Eq. \ref{eqS1}). To optimize the fraction of the muons stopped in the sample, various experimental parameters were optimized, resulting in nearly 50$\%$ sample contribution to the overall signal. 

Above $T_{\rm c}$, in the normal state, the symmetric field distribution can be well described by a single component. The corresponding relaxation rate and internal magnetic field are denoted as $\sigma_{\rm ns}$ and $B_{{\rm int},\rm {s},{\rm ns}}$. $\sigma_{\rm ns}$ is found to be small and temperature independent (dominated by nuclear magnetic moments) above $T_{\rm c}$ and assumed to be constant over the entire temperature range. Below $T_{\rm c}$, in the SC state, the relaxation rate and internal magnetic field are denoted as $\sigma_{\rm SC}$ and $B_{{\rm int},\rm {s},{\rm sc}}$, respectively. The superconducting contribution to the relaxation rate is obtained via $\sigma_\mathrm{SC} = \sqrt{\sigma^{2} - \sigma^{2}_\mathrm{ns}}$.

For a perfect triangular lattice, the relaxation rate is directly linked to the magnetic penetration depth \(\lambda_\text{eff}\) according to the equation\cite{brandt1988flux}:

\begin{equation}
	\begin{aligned}
		\frac{\sigma_\text{SC}(T)}{\gamma_{\mu}}=0.06091\frac{\Phi_{0}}{\lambda_\text{eff}^2(T)}
		\label{eqS12}
	\end{aligned}
\end{equation}

where \(\Phi_{0}\) stands for the magnetic-flux quantum.

${\lambda}$($T$) was calculated within the local (London) approximation (${\lambda}$ ${\gg}$ ${\xi}$) by the following expression \cite{Suter69, Tinkham2004}:
\begin{equation}
\begin{split}
	\frac{\sigma_{\rm SC}(T,\Delta_{0,i})}{\sigma_{\rm SC}(0,\Delta_{0,i})}=
	\\ 1 + \frac{1}{\pi}\int_{0}^{2\pi}\int_{\Delta(_{T,\varphi})}^{\infty}(\frac{\partial f}{\partial E})\frac{EdEd\varphi}{\sqrt{E^2-\Delta_i(T,\varphi)^2}},
    \end{split}
\end{equation}
where $f=[1+\exp(E/k_{\rm B}T)]^{-1}$ is the Fermi function, ${\varphi}$ is the angle along the Fermi surface, and ${\Delta}_{i}(T,{\varphi})={\Delta}_{0,i}{\Gamma}(T/T_{\rm c})g({\varphi}$)
(${\Delta}_{0,i}$ is the maximum gap value at $T=0$). The temperature dependence of the gap is approximated by the expression ${\Gamma}(T/T_{\rm c})=\tanh{\{}1.82[1.018(T_{\rm c}/T-1)]^{0.51}{\}}$,\cite{Carrington205} while $g({\varphi}$) describes the angular dependence of the gap and is replaced by 1 for both a single isotropic gap and two isotropic full gaps, and ${\mid}\cos(2{\varphi}){\mid}$ for a nodal $d$ wave gap~\cite{Guguchia8863}.\\


\section*{Acknowledgments}
This work is based on experiments performed at the Swiss Muon Source (S${\mu}$S) Paul Scherrer Insitute, Villigen, Switzerland. Z.G. acknowledges support from the Swiss National Science Foundation (SNSF) through SNSF Starting Grant (No. TMSGI2${\_}$211750). A.C.S. and S.D.W. gratefully acknowledge support via the UC Santa Barbara NSF Quantum Foundry funded via the Q-AMASE-i program under award DMR-1906325.\\

\section*{Author contributions}
Z.G. conceived, designed and supervised the project. Sample growth: A.C.S. and S.W.. Zero and low-field $\mu$SR experiments: P.K., O.G., J.A.K., T.J.H., H.L.,  and Z.G.. High field $\mu$SR experiments: P.K., J.N.G., O.G., A.D. and Z.G. High pressure $\mu$SR experiments: P.K., S.S.I., J.N.G., O.G., R.K., and Z.G.. Data analysis, figure development and writing of the paper: P.K. and Z.G. All authors discussed the results, interpretation, and conclusion.\\ 

\section*{Data availability}
The data that support the findings of this study are available from the corresponding authors upon request.\\

\section*{Conflict of Interest}
The authors declare no financial/commercial conflict of interest.\\


\bibliography{References}{}

\end{document}